\documentclass[a4paper]{article}

\usepackage{INTERSPEECH2022}

\usepackage{multirow}
\usepackage{tabu}
\usepackage{cite}
\usepackage{multirow,xcolor}

\title{On monoaural speech enhancement for automatic recognition \\of real noisy speech using mixture invariant training}
\name{Jisi Zhang$^1$, C\u{a}t\u{a}lin Zoril\u{a}$^2$, Rama Doddipatla$^2$ and Jon Barker$^1$}
\address{
  $^1$University of Sheffield, Department of Computer Science, Sheffield, UK \\
  $^2$Toshiba Cambridge Research Laboratory, Cambridge, UK}
    \email{\{jzhang132,j.p.barker\}@sheffield.ac.uk,
    \{catalin.zorila,rama.doddipatla\}@crl.toshiba.co.uk}

\begin{document}

\maketitle
\begin{abstract}
In this paper, we explore an improved framework to train a monoaural neural enhancement model for robust speech recognition. The designed training framework extends the existing mixture invariant training criterion to exploit both unpaired clean speech and real noisy data. It is found that the unpaired clean speech is crucial to improve quality of separated speech from real noisy speech. The proposed method also performs remixing of processed and unprocessed signals to alleviate the processing artifacts. Experiments on the single-channel CHiME-3 real test sets show that the proposed method improves significantly in terms of speech recognition performance over the enhancement system trained either on the mismatched simulated data in a supervised fashion or on the matched real data in an unsupervised fashion. Between 16\% and 39\% relative WER reduction has been achieved by the proposed system compared to the unprocessed signal using end-to-end and hybrid acoustic models without retraining on distorted data.

\end{abstract}
\noindent\textbf{Index Terms}: semi-supervised learning, speech enhancement, speech recognition, mixture invariant training

\section{Introduction}
Speech recorded in everyday environments is corrupted by ambient noise and reverberation, which degrade the performance of automatic speech recognition (ASR)~\cite{zhang2018deep}. Speech enhancement aims to recover the intelligibility and perceived quality of noisy speech. In this paper we are addressing the single-channel scenario.
Recently, speech enhancement techniques have made great progress driven by the power of deep learning~\cite{Wang2020ComplexSM}. A typical deep learning based enhancement system is built through supervised training, which trains a network to learn a mapping from noisy to clean speech features~\cite{wang2018supervised}. The supervised learning framework requires paired noisy and clean speech samples, which are generally simulated by mixing plain speech with pre-recorded noise signals~\cite{Thiemann2013TheDE,Gemmeke2017AudioSA}.
However, the enhancement system trained on simulated data under-performs when dealing with unseen real data, due to a distribution mismatch between the simulated and real data.  
This mismatch may cause severe distortion to the processed speech signal and degrade speech recognition performance of an acoustic model that has never seen the artifacts. This mismatch problem cannot be fixed simply by retraining with `real data' as in real scenarios there is generally no access to corresponding clean speech signals. The distortion can be reduced by applying array processing techniques such as beamforming when audios are recorded by multiple microphones~\cite{heymann2016neural}. However, this technique is limited to multiple microphone recordings and cannot be applied to single-channel speech enhancement.


In real situations, where paired noisy and clean signals are not available, we may instead look to use \textit{unpaired} noisy and clean speech data.
Several training strategies have been developed for using such data based on adversarial learning~\cite{meng2018cycle,Yuan2019CycleGANbasedSE} and transfer learning~\cite{Sivaraman2021AdaptingSS,Tzinis2022ContinualSW,Tzinis2022RemixITCS}. For the adversarial training, discriminator networks are used to distinguish the enhanced and noised features from the clean and noisy ones, respectively~\cite{meng2018cycle}. An alternative approach is transfer learning, where a teacher-model is trained on out-of-domain synthetic data to infer pseudo-targets for in-domain noisy data. Then, the estimated clean speech and noise signals are remixed via data augmentation strategies to generate a new set of mixtures to train a student model~\cite{Tzinis2022RemixITCS}. However, when the real recordings are highly mismatched to the simulated data used to train the teacher model, the teacher model may not generate reliable pseudo-targets from the real data to train the student model.

Mixture invariant training (MixIT)~\cite{wisdom2020unsupervised} has been developed to train a speech separation or enhancement system without using clean ground-truth signals. The main idea is that it only uses noisy signals to conduct self-training. A noisy speech signal and a non-speech noise signal are mixed as input to an enhancement model to infer individual sources, which are recombined to reconstruct reference noisy speech and non-speech noise. MixIT has been demonstrated to be effective for reducing noise in a noisy speech signal and has achieved comparable results in terms of signal quality measurement, compared to a fully-supervised system.
In~\cite{Sivaraman2021AdaptingSS}, to adapt a speech separation model to real far-field speech data recorded in meeting scenarios, MixIT and a conventional supervised training framework are combined to exploit both mismatched synthetic data and matched real data. However, the separated signals in~\cite{Sivaraman2021AdaptingSS} have been evaluated subjectively by human listeners and it is still unknown whether the unsupervised model could benefit a speech recognizer with real noisy data.
To the best of our knowledge, MixIT has not been explored to jointly use unpaired clean speech and real noisy data for monoaural speech enhancement.

\begin{figure*}[h]
    \centering
    \includegraphics[width=0.85\textwidth]{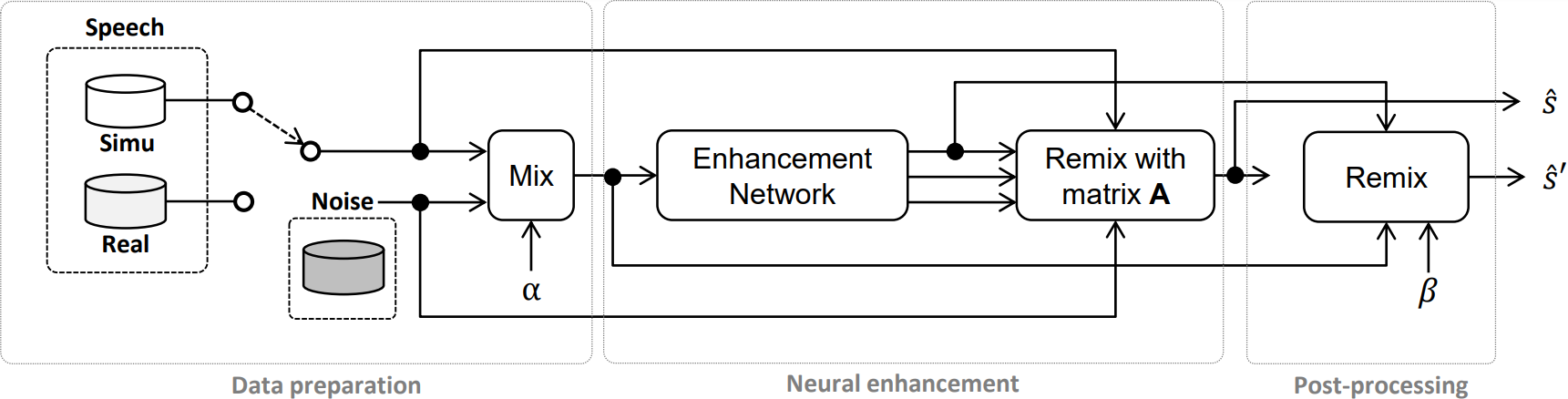}
    \caption{Block diagram of proposed monoaural speech enhancement system.}
    \label{fig:diagram}
\end{figure*}

To address the problem caused by the mismatch between real and simulated scenarios, this work investigates a novel training framework based on MixIT to exploit both unpaired clean speech and real noisy data to train a speech enhancement system for ASR. The designed framework requires out-of-domain clean speech, in-domain background noise, and in-domain noisy speech data. We first show that training on real noisy data alone with MixIT cannot efficiently improve speech recognition performance. It is potentially due to the fact that MixIT is originally designed to train a network to separate independent sources instead of improving the quality of speech. Therefore, we propose to use both out-of-domain clean data and in-domain real data during training. In this case, using clean audio as part of training target encourages the network to enhance the speech from the noisy  signal, while accessing real noisy data helps the model generalise better. We further investigate how the size of the out-of-domain clean speech data and in-domain real data affects the enhancement system. The proposed method is evaluated using the real test sets from CHiME-3. 

The rest of paper is organised as follows. Section~\ref{sec:method} introduces the proposed method. In Section~\ref{sec:exp_setup}, the evaluation and the implementation details are described. Section~\ref{sec:results} presents the results and some discussions. Finally, this paper is concluded in Section~\ref{sec:conclude}.

\section{Method}
\label{sec:method}

The proposed single-channel speech enhancement system used in this investigation has the block diagram depicted in Figure~\ref{fig:diagram}, and consists of three components: data preparation, neural enhancement and post-processing. A brief description of each component is given below, followed by a summary of paper's main contributions.
\subsection{Data preparation}

This stage is responsible for generating training examples for the neural enhancement network.
The monoaural speech signal captured from a single source in a noisy reverberant environment can be modelled in the time domain as
\begin{equation}
    y(t)=h(t)*s(t) + n(t),
\end{equation}
where $s$, $n$ and $h$ are the source, background noise and the room's impulse response, respectively. $*$ denotes the convolution operation. For simplicity, in this work $s$ and $n$ are assumed uncorrelated, and $h$ is time-invariant. Applying the Fourier transform in the equation above yields 
\begin{equation}
    Y(k,f) = H(f)S(k,f) + N(k,f),
\end{equation}
where $k$ and $f$ represent the frame index and the current frequency bin, respectively.

If enough clean speech and noise samples are available, then the enhancement system can be trained in a supervised fashion by mixing speech and noise at various signal-to-noise ratios (SNRs) $\alpha$. Artificially generated room impulse responses can also be convolved with the clean speech signals to improve the simulation.
However, the distribution of simulated data obtained as formerly described cannot fully match the distribution of real recordings. A neural network based enhancement system is sensitive to the mismatch and will cause severe distortions to processed signals when the input signal is mismatched to the distribution of training data.

Mixture invariant training (MixIT,~\cite{wisdom2020unsupervised}) has been recently proposed for speech separation and enhancement when only noisy signals are available. For the speech enhancement task, MixIT randomly draws a noisy speech signal and a non-speech noise signal from an unlabelled dataset, and adds them together to create an artificial mixture. The enhancement model takes the latter noisy signal as input and predicts several sources, which are remixed to reconstruct the initial signals, as further explained in the next section.

\subsection{Neural enhancement network}

The dense U-Net temporal convolutional network proposed by Wang et al.~\cite{Wang2020ComplexSM} is used for the monoaural speech enhancement. The network consists of an encoder-decoder structure, similar to U-Net~\cite{ronneberger2015u}, and temporal convolutional networks (TCNs)~\cite{lea2016temporal}. The encoder and decoder are constructed from densely connected convolutional blocks~\cite{Huang2017DenselyCC}. The TCN is built from $\mathrm{R}$ repetitions of a sub-block which stacks $\mathrm{X}$ dilated 1-dimensional convolutional blocks.
The short-time Fourier transform (STFT) is used to convert waveforms to the time-frequency domain using frame and hop sizes of 32\,$\mathrm{ms}$ and 8\,$\mathrm{ms}$, respectively. The input to the model is the concatenation of the real and imaginary components of the mixture's STFT.

The network is trained in a MixIT-style to optimise a complex spectral mapping (CSM) loss, which has been demonstrated to be beneficial for speech recognition~\cite{Wang2021MultimicrophoneCS}. The CSM loss consists of mean absolute error between real, imaginary and magnitude components of estimated and reference sources. The reference sources are randomly drawn from an unlabelled dataset and consist of a 
noisy speech signal $x_1$ and a non-speech noise signal $x_2$, which are added together to create an artificial mixture. In what follows, $\mathbf{x}=[x_1;x_2]$ is the reference signal used to train the system. Therefore, the training loss is defined as:
\begin{equation}
\vspace{-5pt}
\begin{split}
		\mathcal{L}_{CSM} =  &\min_{\mathbf{A}} \mathcal{L}_{1} [\mathbf{\mathrm{Re}(X)},\mathbf{A}\mathrm{Re}(\hat{\mathbf{S}})] \\
		& + \min_{\mathbf{A}} \mathcal{L}_{1} [\mathbf{\mathrm{Im}(X)},\mathbf{A}\mathrm{Im}(\hat{\mathbf{S}})] \\
		& + \min_{\mathbf{A}} \mathcal{L}_{1} [\mathbf{|X|},\mathbf{A}|\hat{\mathbf{S}}|],
\end{split}
\end{equation}
where $\mathbf{X} \in \mathbb{C}^{2 \times F \times K}$ and $\hat{\mathbf{S}} \in \mathbb{C}^{M \times F \times K}$ denote the complex STFT of the reference and estimated signals, respectively, 
$\mathrm{M}$ is the number of output channels (i.e, predicted sources) of the enhancement system,
$\mathrm{F}$ is the number of frequency bins and $\mathrm{K}$ is the total number of time frames.
$\mathrm{Re(\cdot)}$ and $\mathrm{Im(\cdot)}$ are operators to extract the real and imaginary parts from a complex number, and $|\cdot|$ represents the magnitude value.
The magnitude component is incorporated into the loss function as the real and imaginary parts cannot lead to accurate magnitude estimates~\cite{Wang2020ComplexSM}.
$\mathbf{A} \in \mathbb{B}^{2 \times M}$ is a mixing matrix whose elements along each column sum to 1 assigning each predicted source $\hat{\mathbf{s}}$ to to either $x_1$ or $x_2$ as proposed in~\cite{wisdom2020unsupervised}. To force the model to output denoised speech in the first channel only, $\mathbf{A}$ is constrained such that channel 1 alone, or channels 1 and 2, or channels 1 and 3 are used to reconstruct the original clean or noisy speech signal.
Channel 2 or channel 3 alone or the their sum are employed to reconstruct the non-speech signal.
The $\mathcal{L}_{1}$ loss is defined as:

\begin{equation}
		\mathcal{L}_{1} [\mathbf{X},\mathbf{A}\hat{\mathbf{S}}] =  \sum_{i=1}^{2} \sum_{f=1}^{F} \sum_{k=1}^{K} |\mathbf{X}_{i,f,k} - (\mathbf{A}\hat{\mathbf{S}})_{i,f,k}|.
\end{equation}

\subsection{Post-processing remix}

Speaker reinforcement has been recently proposed in \cite{Zorila_icassp2022} for improving the ASR accuracy of enhanced speech without acoustic model retraining. The enhanced signal is remixed with the unprocessed noisy input at a given SNR $\beta$ (typically 0), which is found to alleviate the processing artifacts.

\subsection{Main contributions}

The main contributions of this paper are as follows. We: (i) combine the dense U-Net temporal convolutional network proposed by Wang et al.~\cite{Wang2020ComplexSM} with the MixIT strategy to process time-frequency signal representations instead of time-domain signals, under a complex spectral mapping loss optimisation; (ii) show that exploiting out-of-domain clean speech as well as in-domain real noisy data during the training of the enhancement network yields significant recognition gains for real test samples; (iii) use the MixIT framework for both types of data instead of switching to supervised training when using out-of-domain clean speech (as described in~\cite{Sivaraman2021AdaptingSS}); (iv) exploit speaker reinforcement post-processing to mask processing artifacts and further improve ASR accuracy.

\section{Experiment Setup}
\label{sec:exp_setup}
\subsection{Data}

The proposed method is evaluated using the CHiME-3 data~\cite{Barker2015TheT}.
The CHiME-3 corpus consists of simulated and real noisy read speech recordings from the WSJ0 SI-84 corpus~\cite{Paul1992TheDF}. The real data are recorded in five environments (booth, bus, cafeteria, street junction and pedestrian area) using six microphones mounted on a tablet and a close-talk microphone.
The simulated data are obtained by convolving the clean WSJ0 SI-84 speech with time-varying filters estimated from the tablet and close-talk microphones, which is then mixed with real background noise from CHiME-3. The training data includes 7,138 simulated and 1,600 recorded utterances, and the test data contains 1,320 recorded utterances. All data used to perform experiments have 16~kHz resolution.

\subsection{Speech enhancement system configuration}

The temporal convolutional network of the enhancement system is made of $\mathrm{R=2}$ repetitions of a sub-block which stacks $\mathrm{X=7}$ dilated 1-dimensional convolutional blocks. The frame and hop sizes for the STFT are 512 samples and 128 samples, respectively.
During training, artificial noisy data are created by mixing clean speech with background noise from CHiME-3 at SNRs uniformly sampled from [-5,\,5] dB. The mixtures are dynamically created at training time.

Enhancement models are trained using the Adam optimizer~\cite{kingma2014adam} with a learning rate of $1e-3$, which is halved every time the validation loss is not reduced in 3 consecutive epochs.
All models are trained with 100 epochs, batch size of 8, and 2 second speech chunks.
For cross-validation, the dev simu set provided in the official CHiME-3 corpus is used, and the enhancement model that achieves the highest SNR improvement is selected as the best model.



\subsection{Speech recognition system configuration}

Two acoustic models are employed to assess the ASR performance. The first one is a Conformer based end-to-end speech recognition system~\cite{gulati2020conformer} (E2E). Its configuration is the same as described in~\cite{Guo2021RecentDO}: $\mathrm{Enc=12}$, $\mathrm{Dec=6}$, $\mathrm{d^{ff}=2048}$, $\mathrm{H=4}$, and $\mathrm{d^{att}=256}$. The training data is made of clean WSJ0 SI-84 data, synthetic noisy speech, and real noisy speech. The training data is further augmented with speed perturbation~\cite{ko2015audio} at ratios 0.9, 1.0, 1.1 and SpecAugment~\cite{park2019specaugment}. The implementation and pre-trained model are available in ESPNet with the link: ``{https://github.com/espnet/espnet/tree/master/egs/chime4/asr1}''.
The second acoustic model (CHiME-3 Original, C3-ORG) is a hybrid system, which has a 14-layer TDNNF~\cite{Povey_is2018_tdnnf} topology and is trained with the standard noisy set from CHiME-3 (all six channels from the real and simulated training sets) and 3-fold speed perturbation. The acoustic features are 40-dimension MFCCs and 100-dimension i-vectors, and the model is trained in Kaldi using the lattice-free MMI criterion. The baseline tri-gram (3G) and an RNN-based language model (LM) are used for decoding.

The fifth channel from the real dev and eval CHiME-3 test sets is employed for the E2E experiments since this channel exhibits the highest SNR among all microphones. To facilitate comparison with other studies from the literature, the official single-channel test list (random channels) from CHiME-3 is employed with the C3-ORG acoustic model.


\section{Experimental results}
\label{sec:results}
We first show the problem caused by mismatch when training a speech enhancement system on simulated data while testing on real noisy data.
The enhancement system is trained on the official CHiME-3 simulated data via supervised learning and tested on the real noisy recordings. Table~\ref{tab:semi_enh} shows that the enhancement model trained on the mismatched data achieved a WER of 21.2\%, which is worse than the result achieved by the unprocessed signal. This performance drop has also been observed in~\cite{Wang2020ComplexSM}. Though the in-domain noise has been used during data simulation, the results show that the training and test conditions are still mismatched.
Next, we investigate the unsupervised enhancement model by accessing only real data, which is labelled as Unsupervised in Table~\ref{tab:semi_enh}. It is observed that the unsupervised system achieves a 21.8\% WER, which is also worse than the result achieved by the unprocessed signal. This is potentially due to the fact that MixIT is originally designed to train a network to separate out each independent sources given a mixture signal, but the quality of speech components degraded by noise cannot be improved by purely unsupervised training.

\begin{table}[t]
\centering
\setlength{\tabcolsep}{4pt}
\caption{ASR performance (WER\%) on CHiME-3 (channel 5) using E2E acoustic model (no post-processing).}
\label{tab:semi_enh}
\begin{tabu}{lcccc}
\hline
\textbf{Method}          & \textbf{Train Data}      & \textbf{Target}      & \textbf{dev real}     & \textbf{eval real} \\ \hline
\rule{0pt}{1em}Unprocessed              & -                        & -                    & 11.1                  & 18.9               \\
{Supervised}             & Simu                     & Clean                & 11.1                  & 21.2               \\
{Unsupervised}           & Real                     & Noisy                & 12.4                  & 21.8               \\ \hline
\rule{0pt}{1em}{Unsupervised}           & Simu\&Real               & Noisy                 & 11.0                  & 18.4               \\
{Proposed}           & Simu\&Real               & Both          & 10.3                   & {\bf 17.4}               \\ \hline
\end{tabu}
\end{table}

Then, the enhancement model is trained with the proposed method using both synthetic data from the third microphone channel and all the real noisy data. 
We also conducted a control experiment with the same augmented training set. The difference during training is that, when using the simulated data, the input to the network is a mixture of one simulated noisy signal and an additional noise signal. The enhancement model is trained to reconstruct the noise signal and the original simulated noisy signal, instead of clean speech.


The bottom part in Table~\ref{tab:semi_enh} shows the results achieved by the control experiment and the proposed method. As shown in the fourth row, the control experiment improves the ASR performance from 21.8\% to 18.4\% by increasing the amount of noisy data, which is better than the result achieved by the unsupervised learning based on only real data, but the gain is still small in comparison with the unprocessed signal. For the proposed method, reconstructing the clean speech reduces the WER to 17.4\%, as shown in the last row. This means the proposed method successfully maps the domain of real noisy speech to the domain of clean speech without accessing ground truth labels. This improvement is potentially due to the fact that optimising the loss between estimated and clean speech forces the enhancement network to reconstruct a less distorted signal.

Next, to investigate the effect of the amount of unpaired clean speech data used for the proposed training framework, more clean speech data is added to increase acoustic diversity. The result in Table~\ref{tab:semi_unpaired_data_effect} shows that using all simulated data (six channels) reduces the WER to 15.8\% compared to the result obtained by only using one channel, 17.4\%, indicating that the proposed method can exploit the variability in the simulated data to benefit the enhancement of real noisy speech. Remixing the enhanced speech with the unprocessed input at $\beta$=0~dB reduces the WER further (last row in Table~\ref{tab:semi_unpaired_data_effect}).


\begin{table}[t]
\centering
\setlength{\tabcolsep}{4pt}
\caption{ASR performance (WER\%) of proposed system on CHiME-3 (channel 5) using E2E acoustic model for various amounts of simulated data.}
\label{tab:semi_unpaired_data_effect}
\begin{tabu}{cccc}
\hline
\textbf{Simu (hrs)}            & \textbf{Real (hrs)}        & \textbf{dev real}     & \textbf{eval real} \\ \hline
\rule{0pt}{1em}0                            & 19                     & 12.4              & 21.8             \\
3                            & 19                     & 10.6              & 18.0             \\
15                         & 19                  & 10.3               & 17.4             \\
39                         & 19                  & 9.7               & 16.8             \\
75                       & 19                 & 9.4               & 15.8            \\ \hline
{\rule{0pt}{1em} + Remix ($\beta=\mathrm{0}$ dB)}               &                    & {\bf 9.3}                   & {\bf 15.1}               \\ \hline
\end{tabu}
\end{table}

\begin{figure}[htb]
    \centering
    \includegraphics[width=0.5\textwidth]{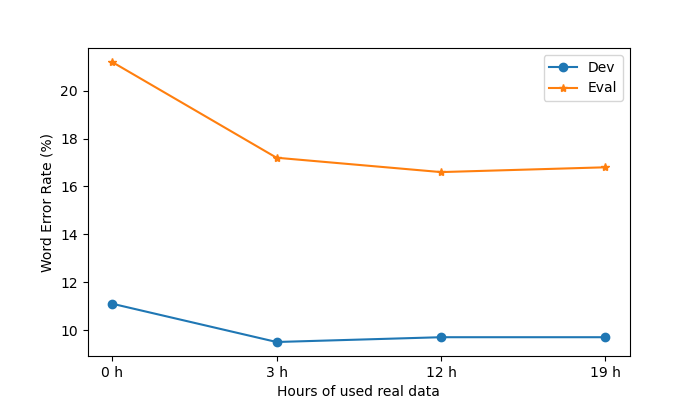}
    \caption{ASR performance (WER\%) of proposed system on CHiME-3 (channel 5) using E2E acoustic model for various amounts of real noisy data. 39 hours of simulated data is included in the training set.}
    \label{fig:effect_real}
\end{figure}

The effect of adding more real noisy training data is also evaluated. 
The result is shown in Figure~\ref{fig:effect_real} and it illustrates that a small amount of matched data substantially improves the speech recognition accuracy, i.e., adding about 3 hours of real training data reduces the WER from 21.2\% to 17.2\%. However, further increasing the amount of real data does not yield more benefit.

The ASR results with the C3-ORG acoustic model are depicted in Table~\ref{tab:comparison}. 
Some top performances reported in the literature for the single-channel CHiME-3 track are also provided for comparison purposes. The proposed method yielded 28\% and 39\% WER reduction compared with the unprocessed signal obtained by the same Kaldi baseline for the real development and evaluation sets, respectively. Without using LSTM language model rescoring and speaker adaptation techniques, the proposed system achieves comparable result with the current state-of-the-art system on the real evaluation set.

For the future work, we would like to retrain the acoustic model with an enhanced training set processed by the proposed enhancement model to further reduce the mismatch. We will also conduct the proposed self-training with larger clean speech databases such as LibriSpeech~\cite{Panayotov2015LibrispeechAA} to further improve the enhancement model.

\begin{table}[t]
\centering
\setlength{\tabcolsep}{4pt}
\caption{ASR performance (WER\%) of proposed system on CHiME-3 (official 1-channel real test sets) using C3-ORG acoustic model.}
\label{tab:comparison}
\begin{tabu}{lcc}
\hline
\textbf{Model}                                                              & \textbf{dev real}     & \textbf{eval real} \\ \hline
{\rule{0pt}{1em}Kaldi baseline~\cite{chen2018building}}  (LSTM LM)                                    & 5.6                   & 11.4          \\
{Du et al.~\cite{du2016ustc}}  (LSTM LM)                                             & 4.5                   & 9.2           \\
{Kinoshita et al.~\cite{Kinoshita_icassp20a}} (RNN LM)                              &    -           &  8.3 \\
{Wang et al.~\cite{Wang2020ComplexSM} (RNN LM)}             & 4.8                   & 8.4          \\
\rowfont{\color{gray}}{{~~}+ LSTM LM}                                                  & 4.1                   & 8.1          \\
\rowfont{\color{gray}}{{~~~~}+ iterative speaker adaptation}                             & 3.5                   & 6.8          \\ \hline
{\rule{0pt}{1em}Proposed (3G LM)}                                                          & 6.4                   & 10.7         \\ 
{{~~}+ Remix ($\beta=\mathrm{10}$ dB)}                                           & 6.1                   & 9.9         \\ 
{{~~~~}+ RNN LM}                                                                  & {\bf 4.0}                   & {\bf 6.9}         \\ \hline
\end{tabu}
\end{table}

\vspace{-5pt}
\section{Conclusions}
\label{sec:conclude}
In this paper, we have developed an improved framework for monoaural neural enhancement, which: (i) combines the mixture invariant training criterion with a frequency domain enhancement network, (ii) exploits unpaired clean speech and real noisy data, and (iii) performs remixing of processed and unprocessed signals.
It has been shown that exploiting both unpaired clean speech and matched real noisy data can improve a single-channel speech enhancement system built using the mixture invariant training framework. The proposed method has achieved state-of-the-art performance on the official single-channel CHiME-3 track without retraining the speech recogniser. Using an end-to-end acoustic model, the processed speech yielded 16\% and 20\% WER reduction compared with the unprocessed signal for the real development and evaluation sets, respectively. With a more powerful hybrid acoustic model, the processed speech yielded 28\% and 39\% WER reduction compared with the unprocessed signal for the real development and evaluation sets, respectively.



\bibliographystyle{IEEEtran}

\bibliography{mybib}

\end{document}